# HOT-WIRE BASED ESTIMATION OF PRESSURE FLUCTUATIONS IN THE NEAR FIELD OF A JET IN THE PRESENCE OF A CO-FLOW

## O.P. Bychkov and G.A. Faranosov*

*Central Aerohydrodynamic Institute, Moscow Research Complex, Moscow, 105005 Russia*

*\*e-mail: georgefalt@rambler.ru*

It is shown that the velocity fluctuation spectra measured using a hot-wire in the potential flow region of a turbulent jet near field in the presence of a co-flow can be converted into the spectra of pressure fluctuations. The proposed conversion method is based on the fact that the structure of instability waves, which make a decisive contribution to the jet near-field fluctuations, resembles homogeneous one-dimensional waves, which makes it possible to locally link the pressure fluctuations and the fluctuations of the streamwise velocity component measured by a hot-wire.

## INTRODUCTION

A jet issuing out of a circular nozzle at high Reynolds numbers is a significant source of aviation community noise [1, 2], as well as a source of significant pressure pulsation loads for airframe elements [3, 4]. The success of solving applied problems, such as reducing jet noise or related structural loads, is closely related to the progress in the theoretical description of non-stationary processes taking place in the jet, which in turn reflects the depth of our understanding of their physical nature and allows us to propose more effective ways in solving engineering tasks.

To date, a complete dynamic model of a turbulent jet as an object of fluid dynamics has not yet been developed; in particular, there is no commonly accepted point of view on the nature of noise generation by such a flow [5-10]. From the viewpoint of the jet near-field fluctuations, the situation is clearer. It is believed that, for the spectral maximum, the fluctuations are mainly associated with the development of coherent structures in the jet mixing layer that arise due to the Kelvin-Helmholtz instability and have the form of wave packets with different azimuthal numbers [10-13]. It is disturbances of this type that determine the maximum levels of fluctuations in the near hydrodynamic field of the jet (outside the mixing layer) and are responsible for a significant additional noise associated with their scattering at the wing trailing edge in the case of its close location to the nozzle, as is often the case for modern aircraft [3, 14-17]. In [13, 18, 19], the possibility of active control of instability waves was studied in order to reduce their intensity in the near field and accordingly attenuate the noise of the interaction of the jet closely integrated with the wing [20, 21]. In addition, in high-speed jets, instability waves can themselves become noticeable sound sources even in the absence of nearby scattering surfaces [7, 9, 10].

To build near-field models that allow, for example, to evaluate the spectral composition and fluctuation levels, and to develop effective systems for instability wave suppression, sufficiently detailed information about their characteristics (azimuthal content, spatial structure, phase velocities, etc.) is required. In most studies, such information is obtained from direct measurements by microphones located near the outer boundary of the mixing layer [4, 11, 13, 17, 19, 22, 23], less often – using measurements of the velocity field with a non–invasive particle image velocimetry (PIV) technique [12]. Moreover, measurements of this kind are usually [4, 11-13, 17, 19, 22] carried out under static conditions, when the jet issues into a stationary medium, while from a practical point of view, the case of the jet issuing into a co-flow is of considerable interest, because this is exactly the situation that is realized in flight conditions. Measurements of the near-field pressure fluctuations for a free jet in the presence of the co-flow are difficult, because they require placing a microphone array into the co-flow. This inevitably leads to the occurrence of parasitic disturbances, and, in addition, the microphones for such measurements must be equipped with special nose cones and be oriented in a certain way with respect to the incoming flow [23]. A hot-wire provides a natural, affordable and relatively non-invasive way of measuring fluctuations in the flow. However, the problem of converting velocity fluctuations into pressure ones generally cannot be solved within the framework of single-point measurements. To solve it, knowledge of the velocity fluctuation field in a certain area is required in order to calculate the pressure field based on solving the Euler or Navier-Stokes equations. In general, modern PIV systems for non-intrusive measurements of the velocity field make it possible to solve this problem, at least in an incompressible approximation, when the problem reduced to solving a Poisson type equation [24]. However, such measurements are rather complex and require time-consuming postprocessing [25-27], in addition, PIV is inferior to hot-wire measurements in terms of the spectral resolution. Nevertheless, in some special cases, single-point velocity measurements, easily implemented using a hot-wire, may be sufficient to assess pressure pulsations. Let us consider in more detail in which zones of the fluctuation field produced by a turbulent jet such cases are realized.

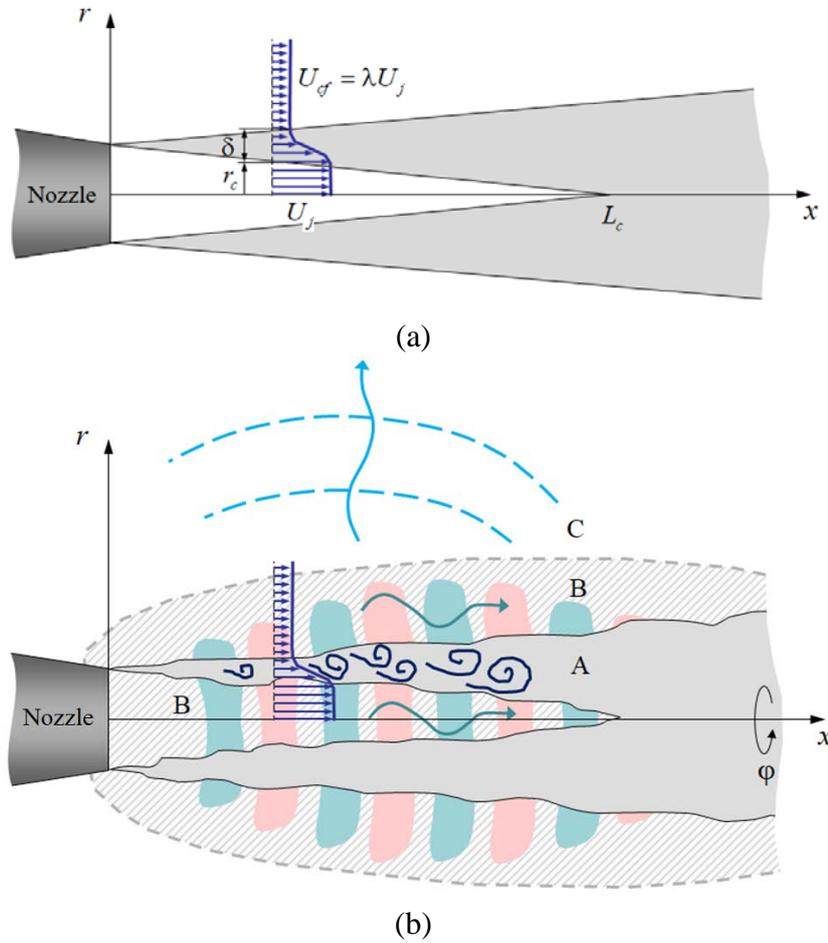

(a)

(b)

Fig. 1. Schematic representation of the mean flow (a) and the structure of fluctuations (b) produced by a turbulent jet issuing from a round nozzle in the presence of a co-flow.

The main object of the present study is a turbulent jet issuing from a round nozzle at a velocity $U_j$ into a uniform co-flow with velocity $U_{cf}$. We will also introduce the co-flow factor $\lambda = U_{cf} / U_j$. Hereafter we assume $\lambda < 1$ which is typical for aircraft jet engine.

Figure 1 shows an illustration of the typical structure of the mean velocity field and the fluctuation field in the symmetry plane of the nozzle [28]. The radius of the potential core is indicated as $r_c$, $\delta$ is the thickness of the mixing layer, $L_c$ is the length of the potential core. In the initial part of the jet, the radial profile of the mean velocity is constant in the potential core $r < r_c$ (here the flow velocity is equal to the velocity of the jet at the nozzle exit plane) and outside the jet $r > r_c + \delta$. The most intense fluctuations associated with turbulent gas motion are observed in the mixing layer (zone A in Fig. 1b). These fluctuations are nonlinear and are described by a complete system of Navier-Stokes equations. In this case, the mean mixing layer is unstable according to the Kelvin-Helmholtz mechanism and causes the development of linear large-scale perturbations – instability waves developing from the nozzle edge downstream in the form of wave packets [7, 9-11]. In the mixing layer, these perturbations are small compared to the small-scale vortex fluctuations and can only be detected by special data processing techniques [12, 29, 30]. However, outside the mixing layer (in the zone of potential mean flow), the intensity of the small-scale fluctuations decreases significantly faster with distance than the intensity of those associated with instability waves (Fig. 1b, zone B). In the potential part of the jet near field, in the vicinity of the mixing layer boundaries, it is the pulsations associated with instability waves that become dominant [9, 11, 17, 31]. Further away from the jet (zone C, fig. 1b) linear acoustic disturbances come to the fore, which, although weak, have the slowest decrease in intensity with distance ($\sim 1/r^2$) compared to the vortex fluctuations and instability waves.

The present study is devoted to the development of a method that allows estimation of the near-field pressure fluctuation spectra (zone B in Fig. 1b) in the presence of the co-flow using a single hot-wire, which is of considerable practical interest. The rest of the paper is organized as follows. Section 1 presents the derivation of simplified relations linking pressure and velocity fluctuations in zone B. Section 2 is devoted to the analysis of the experimental data in the jet near field and the development of a procedure for estimating the convective velocity of disturbances in the presence of the co-flow. Section 3 presents the results of validation of the proposed technique. In conclusion, the main results of the work are formulated.

## 1. THE RELATIONSHIP BETWEEN PRESSURE AND VELOCITY FLUCTUATIONS IN THE POTENTIAL REGION OF THE JET NEAR FIELD

Let us consider in more detail the structure of the fluctuations in zone B in the region of the initial part of the jet ($0 \leq x \leq L_c$). It is in this zone of the jet that airframe elements are often located, which leads to the emission of additional noise [15-17]. It is important that in many cases the jet does not impinge on the airframe elements, i.e. it is aeroacoustic rather than aerodynamic interaction. In this area, pressure perturbations (which are usually measured in experiments) associated with instability wave packets, can be approximated in a cylindrical coordinate system $(x, r, \varphi)$ at each frequency in the form of a superposition of azimuthal harmonics $\hat{p}_n(x, r, \omega)e^{-i\omega(t-x/U_n)+in\varphi}$, where $n$ is the order of the azimuthal mode, $\hat{p}_n$ is a rather slowly changing smooth function of $x$, which is the envelope for a longitudinal wave $e^{i\omega x/U_n}$, and $U_n$ is the phase velocity of the disturbances in the wave [17].

Assume also that the jet is isobaric and isothermal, i.e. the average values of pressure and density (and therefore temperature and speed of sound) are the same throughout the medium, which is quite true for subsonic unheated jets typically used in laboratory experiments. The velocity $U_n$ and shape $\hat{p}_n$ of the wave packet included in (1) can be found from solving the eigenvalue problem for a slowly changing mixing layer [9, 11, 31, 32] or (under static conditions) measured directly in an experiment using near-field microphone arrays [11, 13, 17].

Due to the fact that the mean flow in the jet is characterized by a relatively slow change in the streamwise direction, in the main approximation, the function $\hat{p}_n$ can be considered independent of $x$, and then the perturbations (1) will take the form of plane (in the streamwise direction) waves of the form

$$p_n(r,\omega)e^{-i\omega t - i\alpha_n x + in\varphi},\tag{1}$$

where $\alpha_n = -\omega / U_n$ is the wave number for a given azimuthal mode.

In the potential region (zones B, C in Fig. 1b), pressure perturbations, homogeneous in $x$, satisfy the following equations [13, 17]

$$\left(\frac{\partial^2}{\partial r^2} + \frac{1}{r}\frac{\partial}{\partial r} - \frac{n^2}{r^2} - \beta_j^2\right)p_n = 0, \qquad r \leq r_c,\tag{2}$$

$$\left(\frac{\partial^2}{\partial r^2} + \frac{1}{r}\frac{\partial}{\partial r} - \frac{n^2}{r^2} - \beta_{cf}^2\right)p_n = 0, \qquad r \geq r_c + \delta,\tag{3}$$

where $\beta_j = \sqrt{\alpha_n^2 - \left(k - M_j\alpha_n\right)^2}$, $\beta_{cf} = \sqrt{\alpha_n^2 - \left(k - M_{cf}\alpha_n\right)^2}$, $k = \omega / c$, $M_j = U_j / c$, $M_{cf} = U_{cf} / c$, $c$ is the speed of sound. The multivalued functions $\beta_j$ and $\beta_{cf}$ are defined in accordance, for example, with [33] so that for subsonic flows ($U_n < U_j < c$) $\beta_j > 0$ and $\beta_{cf} > 0$. The solutions of equations (2) and (3) satisfying the finitness condition for $r \to 0$ and radiation condition at $r \to \infty$ have the form

$$p_n = A_n I_n(\beta_j r), \quad r \leq r_c,\tag{4}$$

$$p_n = B_n K_n(\beta_{cf} r), \quad r \geq r_c + \delta,\tag{5}$$

where $I_n$ and $K_n$ are modified Bessel functions of order $n$, of the first and second kind, respectively. Solutions (4) and (5) are essentially components of the solution of the Rayleigh type equation [9, 11, 31] for an axisymmetric mixing layer, which turn into equations (2) and (3) in the corresponding areas of the potential flow. By solving such an equation, it is possible to determine a solution for all $r$ and find, among other things, the relationship of amplitudes $A_n$ and $B_n$ in (4) and (5); however, for the purposes of this work this is not necessary.

Pressure disturbances in the potential region of the near field, given by expressions (1), (4), (5), can be related with the fluctuations of the streamwise velocity component, which have the form $u_n(r,\omega)e^{-i\omega t - i\alpha_n x + in\varphi}$ due to the linearity of the problem. Using the momentum conservation equation in projection on the $x$ axis, one obtains

$$p_n = \rho_0(U_n - U_j)u_n, \quad r \leq r_c, \tag{6}$$

$$p_n = \rho_0(U_n - U_{cf})u_n, \quad r \geq r_c + \delta. \tag{7}$$

Expressions (6), (7) allow us to convert the fluctuations of the streamwise velocity component, measured by a hot-wire, into the pressure fluctuations. An expression for the velocity fluctuations in the jet core, similar to (6), and taking into account the streamwise inhomogeneity of the wave packet, was obtained in [31]. Since it may be problematic to perform direct hot-wire measurements outside the jet under static conditions ($U_{cf} = 0$), the eigen-value problem for a mixing layer with a given mean velocity profile (corresponding to the given section $x$) was solved [31]. The obtained solution made it possible to recalculate the data from the jet axis to the outer region ($r \geq r_c + \delta$) and, hence, to estimate pressure fluctuations in this region. However, in the presence of a co-flow, direct measurements of velocity pulsations are possible in all regions, including the region outside the jet, which is of main practical interest. Thus, according to the simple model described above, for a jet in a co-flow, a hot-wire can be used to directly estimate the spectrum of pressure fluctuations at any point in the region B (Fig. 1b). With this approach, as follows from (7), in addition to the velocity disturbances themselves, it is necessary to know their convective velocity $U_n$ for each significant azimuthal mode $n$. Generally speaking, the determination of these parameters requires multi-channel hot-wire measurements or measurements using time-resolved PIV method. However, the analysis of the experimental data obtained for the near-field fluctuations under static conditions allowed us to construct a fairly simple procedure for a semi-empirical estimation of the convective velocity $U_n$ of the disturbances in the presence of the co-flow based on the results of the direct measurements of $U_n$ in static conditions, as is shown below.

## 2. ANALYSIS OF MEASUREMENT DATA IN THE JET NEAR FIELD

For the purposes of this work, an experimental database obtained in the anechoic chamber AC-2 TsAGI is used. AC-2 chamber is an open-circuit wind tunnel with an open test section designed to study the noise of jets, airframe elements and propellers in static conditions and in the presence of a co-flow simulating flight conditions at takeoff and landing regimes. The test section of AC-2 is an anechoic chamber of dimensions 9.6 m × 5.5 m × 4.0 m. The walls of the chamber are lined with sound-absorbing wedges made of basalt fiber, providing a deviation from free field conditions of no more than 1 dB in the operating frequency range 160-20000 Hz for sound waves with an intensity of up to 160 dB.

The data analyzed in the current work were obtained for a subsonic jet issuing from a round profiled nozzle with a diameter of $D = 40$ mm [13, 17, 34]. The acoustic jet Mach number, based on the speed of sound in the environment, is $M_j = 0.4$ ($V_j \approx 137$ m/s). The Reynolds number, based on the nozzle diameter and the jet velocity, was about $4 \cdot 10^5$. The jet was studied both in static and flight conditions at co-flow factors $\lambda = 0, 0.15, 0.23, 0.3$.

Two types of measurements were carried out: the measurements of pressure fluctuations by a multi-microphone near-field array [13, 17], and the hot-wire measurements of velocity on the jet axis and in several cross sections [34]. Microphone array measurements were carried out only for static conditions ($\lambda = 0$), while hot-wire measurements were performed for all of the above values of $\lambda$. Note that similar measurements were carried out for jets with higher flow velocities (up to $M_j = 0.7$), but they are not considered in this paper.

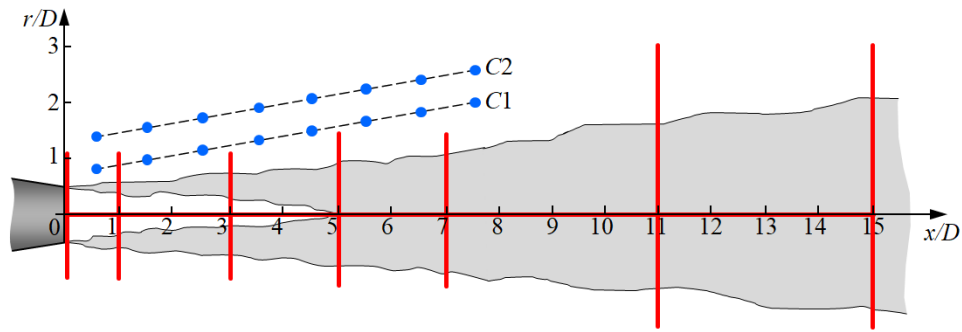

Fig. 2. Location of measurement points. The blue dots are the location of the microphones (shown for one azimuth angle) [13, 17], the red lines show the hot-wire traverses [34].

The near-field array contained 30 microphones and allowed simultaneous measurements of pressure fluctuations in five cross sections, with six microphones evenly distributed around the circumference in each section. Thus, this system allowed resolving the dominant azimuthal modes of the order $n = 0, 1, 2$. The radial position of the microphones in the array were adjusted so as they located on conical surfaces (C1, C2, Fig. 2) with a half angle of 10.5°. The longitudinal coordinates of the cross sections swept the area from $x/D = 0.5$ to *x/D* = 7.5 with a step of Δ*x/D* = 1 (Fig. 2). Bruel&Kjaer ¼" microphones (type 4957, frequency range 50-10000 Hz) were used for the measurements.

Hot-wire measurements were carried out along the jet axis with a step of Δ*x/D* = 0.5, as well as in cross sections with coordinates *x/D* = 0.075, 1, 3, 5, 7, 11, 15 in increments of Δ*r/D* = 0.025, 0.025, 0.025, 0.05, 0.05, 0.1, 0.1, respectively. The sketch of the hot-wire traverses is shown in Fig. 2. The Dantec 55P01 sensor was used, the sampling rate was 80 kHz. The wire of the probe was set perpendicular to the mean flow and thus recorded mainly the streamwise velocity component, which can be represented as a time-averaged value and fluctuations: $U = U_a + u$. A more detailed description of the measurement points and data processing methods are given in [13, 17, 34].

Let us first analyze the results of the pressure fluctuation measurements by the microphone array. From the viewpoint of the practical application of Eqs. (6) and (7), we are interested primarily in the azimuthal content of the fluctuations and their convective velocity. In [17], it was found that in the near field of the jet (region B in Fig. 1b), at low and medium frequencies, hydrodynamic disturbances corresponding to the first two azimuthal modes *n* = 0, 1 dominate, and the shapes of the spectra of these modes as well as their convective velocities differ slightly from each other (similar conclusions were made in [35] based on numerical simulation). Thus, at least for static conditions, it is possible to introduce a single convection velocity $U_c = U_0 \approx U_1$ of the hydrodynamic disturbances discussed in section 1. It was also shown in [17] that for subsonic jets in a given section of the initial part of the jet, the dependence of the normalized convection velocity $\bar{U}_c = U_c / U_j$ on the Strouhal number $St = f D / U_j$ is universal for various jets, i.e. it does not depend on the jet velocity $U_j$. Thus, for static conditions, $\bar{U}_c$ is a function of only two dimensionless parameters: $\bar{U}_c = \hat{g}(\bar{x}, St)$.

Figure 3 shows the convective velocities for the mode *n* = 0, as a function of the Strouhal number, determined in the experiment based on the phase of the cross-correlation function of the signals in the neighboring sections (see also [17]). When moving downstream from the nozzle exit, the spectra of the hydrodynamic fluctuations shifts to lower-frequencies [17], therefore, at certain Strouhal numbers (decreasing with increasing $\bar{x}$), the contribution of the acoustic fluctuations becomes dominant, and the phase velocity significantly increases (it becomes greater or equal to the speed of sound). Such a jump in the phase velocity is clearly visible in Fig. 4 for the cross sections $\bar{x} \geq 3$. In the cross section $\bar{x} = 2$, the phase velocity of the hydrodynamic disturbances is determined for a wide range of Strouhal numbers up to $St = 1$. The values of $\overline{U}_c$ determined from the experiment allow us to propose the following empirical approximation of the function $\hat{g}$

$$\overline{U}_c = \hat{g}(\overline{x}, \text{St}) \approx U_m(\overline{x})(1 - \exp(-\text{St}/\text{St}_m(\overline{x}))),$$  (8)

where $U_m(\overline{x}) = 0.1\overline{x} + 0.44$, $St_m(\overline{x}) = 0.01\overline{x} + 0.105$. Figure 3 shows that the model (8) accurately approximates the results of the experiment under static conditions in the range of Strouhal numbers and coordinate $\overline{x}$ of interest.

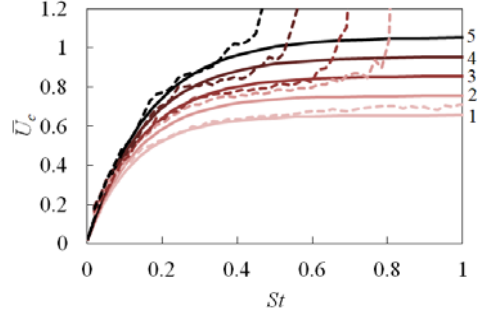

Fig. 3. Dependence of the phase velocity of axisymmetric perturbations ($n = 0$) on the Strouhal number in various cross sections: $1 - \overline{x} = 2$; $2 - \overline{x} = 3$; $3 - \overline{x} = 4$; $4 - \overline{x} = 5$; $5 - \overline{x} = 6$. Dotted lines represent experimental data, solid lines correspond the approximation (8).

An independent verification of the model (8) can be performed using expression (5) describing the behavior of the intensity of the hydrodynamic fluctuations along the radial direction in the region external to the jet. In accordance with (5), the spectrum of these fluctuations $L_{p,n}(St, \overline{x}, \overline{r_1})$ for a given azimuthal mode $n$, measured at the point $(\overline{x}, \overline{r_1})$, where $\overline{r} = r/D$, is related to the corresponding spectrum at the point $(\overline{x}, \overline{r_2})$ by

$$L_{p,n}(\text{St}, \overline{x}, \overline{r_2}) = L_{p,n}(St, \overline{x}, \overline{r_1}) \left| \frac{K_n(\beta_{cf} r_2)}{K_n(\beta_{cf} r_1)} \right|^2.$$  (9)

Figure 4a shows pressure spectra for the axisymmetric mode ($n = 0$), measured on the surface C1 in various sections $\overline{x}$ and recalculated to the corresponding points of the surface C2 (larger radius, Fig. 2) using (9) together with the model (8). For comparison, the spectra directly measured on C2 are given. It can be seen that the model (8)-(9) reproduces quite accurately the behavior of the spectrum of the low-frequency fluctuations as a function of $\overline{r}$. As noted above, the main contribution to the near-field disturbances is made by modes $n = 0$, 1 having similar convection velocities. In addition, an estimate $K_1(\beta_{cf} r) \approx K_0(\beta_{cf} r)$ is valid for the considered values $U_j$, $St$, and $\overline{r}$, which makes it possible to use expression (9) in the form

$$L_p(\text{St},\overline{x},\overline{r_2}) \;\approx L_p(\text{St},\overline{x},\overline{r_1}) \left| \frac{K_0(\beta_{cf} r_2)}{K_0(\beta_{cf} r_1)} \right|^2 \qquad (10)$$

for radial scaling of the total near-field fluctuation spectra, where the value of $\beta_{cf}$ is calculated for the wavenumber $\alpha_n = -\omega / (U_j \hat{g}(\overline{x}, St))$ using the model (8). In Fig. 4b, using the example of recalculation of the pressure spectra from the surface C1 to the surface C2, it is shown that the simplified model (10) is suitable for radial scaling of the spectra, and expression (8), at least for static conditions, correctly describes the convection velocity of the near-field linear hydrodynamic disturbances as a function of $\overline{x}$ and $St$.

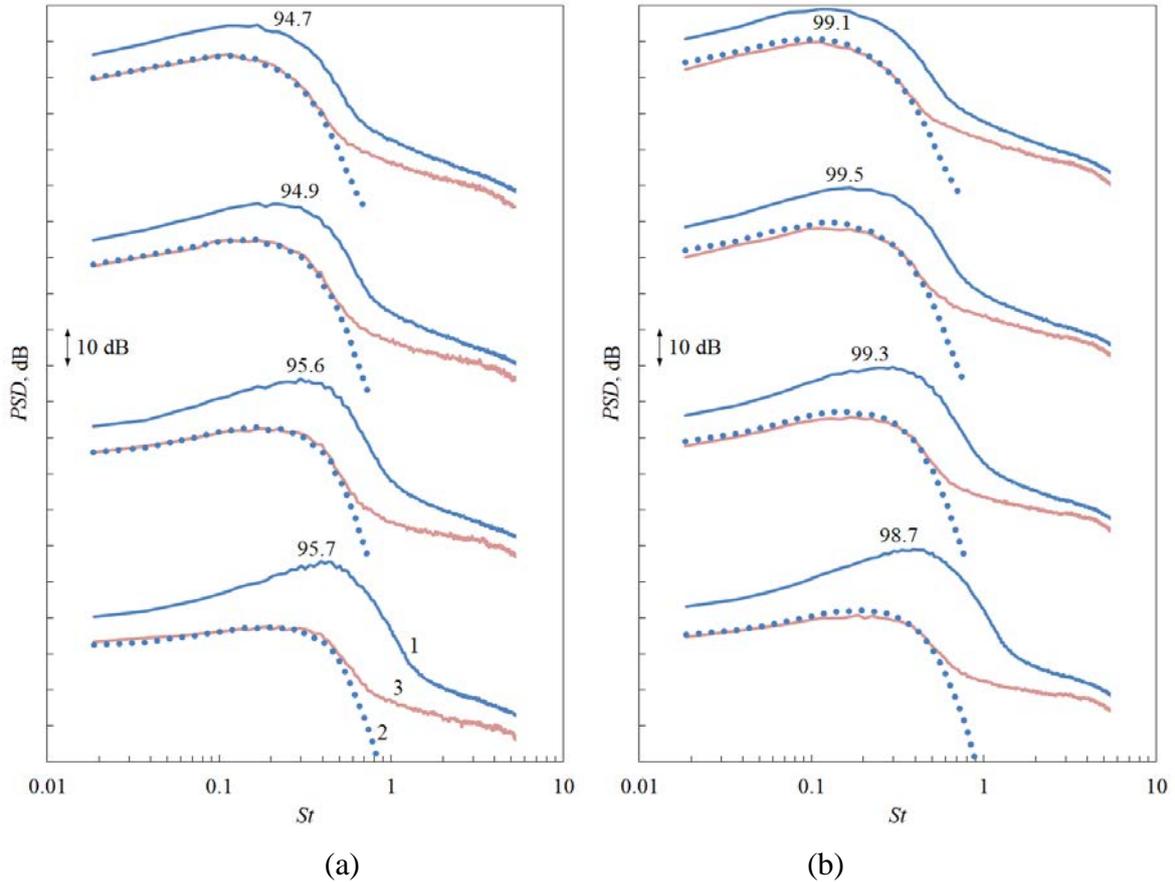

(a)                                    (b)

Fig. 4. Radial scaling of pressure spectra: (a) – spectra of mode $n = 0$; (b) – spectra of total fluctuations. The data for sections $\overline{x}$ = 1.5, 2.5, 3.5, 4.5 (bottom-up) are given. 1 – spectra measured on the surface C1, 2 – spectra measured on the surface C2, 3 – spectra measured on the surface C1 and converted to those on C2. Spectra are plotted in relative units, the maximum level is indicated for each set.

Let us consider the possibility of obtaining an expression, based on (8), for the convective speed of the disturbances in the presence of the co-flow. From the point of view of the stability problem, the convection speed of the disturbances (as well as their increment, if we are interested in the longitudinal structure of the wave packet) is determined by the shape of the velocity profile in the mixing layer [31, 32], which varies along the jet as $\bar{x}$ increases. This shape, in turn, in the initial part of the jet is characterized by two dimensionless parameters: the radius of the potential core $\bar{r}_c = r_c / D$ and the thickness of the mixing layer $\bar{\delta} = \delta / D$. If the width of the mixing layer in a given section of the initial part of the jet is defined as the radial distance from the boundary of the potential core to the point at which the mean velocity drops by a factor of $e$, then the mean velocity profile at $\bar{r} \geq \bar{r}_c$ can be described as $\bar{U}_a = \exp(-(\bar{r} - \bar{r}_c)^2 / \bar{\delta}^2)$, where the parameters $\bar{r}_c$ and $\bar{\delta}$ depend on $\bar{x}$ [11, 28]. This dependence reflects the similarity of the dimensionless velocity profiles in the cross sections of various subsonic jets [28]. This dependence is easily extended to the case with the co-flow, and the transverse profile of the average velocity in the initial part is represented in dimensionless form as [34]

$$\bar{U}(\bar{x}, \bar{r}) = \begin{cases} 1, & \text{for } \bar{r} < \bar{r}_c; \\ \lambda + (1-\lambda)\exp\left(-\left(\dfrac{\bar{r} - \bar{r}_c}{\bar{\delta}}\right)^2\right), & \text{for } \bar{r} \geq \bar{r}_c. \end{cases} \tag{11}$$

In this case, the values $\bar{r}_c$ and $\bar{\delta}$ will depend not only on $\bar{x}$, but also on the co-flow factor $\lambda$. Note that in (11) the boundary layer on the outer part of the nozzle is not considered. The influence of the external boundary layer can be noticeable near the nozzle exit at sufficiently large values of the co-flow factor. However, for the purposes of this work, as will be seen below, a simplified expression (11) is sufficient. A version of the model considering the boundary layer on the nozzle is given in [34].

Experiments show that the main parameter that determines the similarity of jets in the initial section for different $\lambda$, is the length of the potential core $\bar{L}_c$, while $\bar{r}_c$ and $\bar{\delta}$ depend almost linearly on $\bar{L}_c$ [28]. For the co-flow factors under consideration, the values $\bar{L}_c$ measured at different values of $\lambda$ can be approximated by the dependence

$$\bar{L}_c \approx 9.61\lambda + 4.78, \tag{12}$$

while $\bar{r}_c$ and $\bar{\delta}$ can be approximated by the following linear functions

$$\bar{r}_c \approx 0.5(1 - \bar{x} / \bar{L}_c), \ \bar{\delta} \approx 0.6\bar{x} / \bar{L}_c. \tag{13}$$

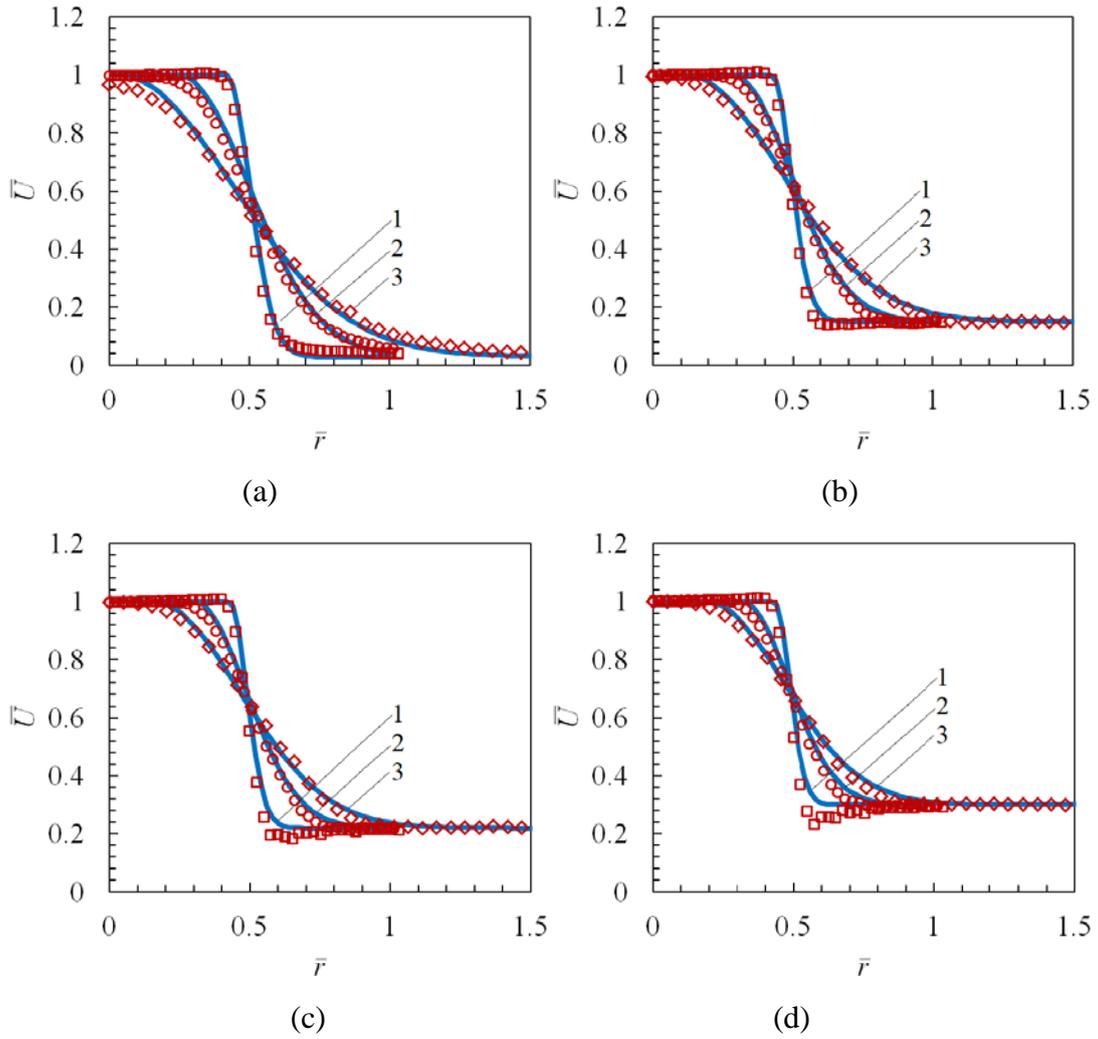

(a)                    (b)

(c)                    (d)

Fig. 5. Transverse profiles of average speed for a jet $M_j = 0.4$ with different co-flow factors: (a) – $\lambda = 0$; (b) – $0.15$; (c) – $0.23$; (d) – $0.3$. Symbols represent experiment, lines denote (11)-(13) model. Profiles shown for three sections of the initial part: $1 - \bar{x} = 1$; $2 - \bar{x} = 3$; $3 - \bar{x} = 5$.

A comparison of the experimentally measured mean velocity profiles and those calculated using model (11)-(13) is shown in Fig. 5. As can be seen, the model quite accurately describes the behavior of the mean velocity profile in the initial section of the jet for various $\lambda$. At $\bar{x} = 1$, the deviation of the model from the experiment is visible as $\lambda$ increases (see Fig. 5c, d), which is associated with the above-mentioned influence of the boundary layer on the outer part of the nozzle.

The validity of model (11)-(13) allows us to conclude that for two subsonic jets, with different velocities and co-flow factors, the velocity profiles in the mixing layer in the corresponding cross sections with respect to the jet potential core lengths will have the same shape (within the accuracy of the semi-empirical model, of course). This statement allows us to propose the following model problem for estimating the relationship between the convection speeds of the disturbances in a jet with a co-flow and in a jet flowing into a stationary medium.

Let us assume, as is often done in analytical models [9, 11, 31, 32], that the expansion rate of the mixing layer in the jet is small. In this case, in each section of the jet the characteristics of the disturbances associated with the instability of the mixing layer (including the phase velocity of the perturbations of interest) can be determined in the main approximation within the framework of a locally parallel model. In this model, the inhomogeneity of the mean field in the axial direction is neglected. From this point of view, let us consider the simplest model of a jet with a co-flow. Let $U(r)$ be the transverse velocity profile of a parallel axisymmetric flow, for which $U(r) = U_j = \text{const}$ for $r < r_c$ and also $U(r) \to U_{cf} = \text{const} < U_j$ for $r \to \infty$ (Fig. 6a).

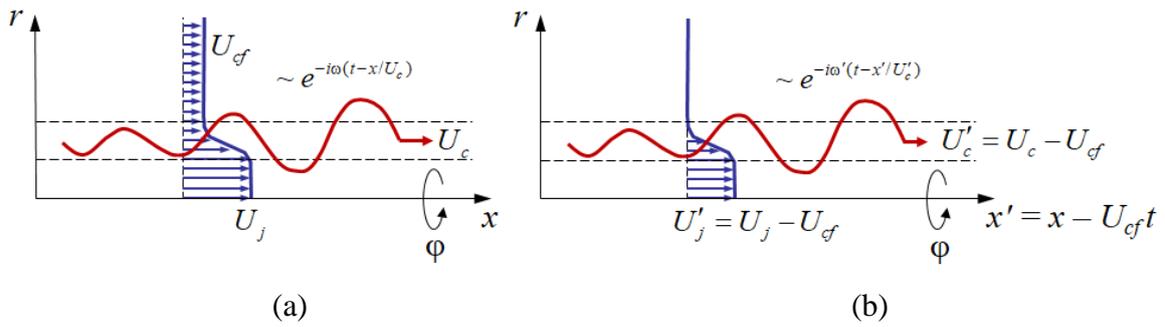

(a)                                          (b)

Fig. 6. Transverse velocity profiles for a homogeneous problem: (a) – in the original coordinate system $(x, r)$ in the presence of a co-flow $U_{cf}$; (b) – in the coordinate system $(x', r)$, associated with co-flow.

Let a disturbance, which is a plane wave in the $x$ direction of the form $e^{-i\omega(t - x/U_c)}$, propagate on the mean flow $U(r)$, where $\omega$ is a given frequency, and $U_c$ is the speed of wave propagation. In particular, we can assume that disturbances in the form of instability wave packets have similar structure if, within the framework of the locally parallel approximation, the axial evolution of the packet's envelope in the jet is not taken into account. Let us move to the reference frame $(x', r)$ with $x' = x - U_{cf}t$, associated with the co-flow (Fig. 6b). We will denote by a prime the parameters related to the reference frame $(x', r)$. Thus, the average flow velocity $U'(r)$ in the new coordinate system will correspond to the model of a jet flowing into a stationary medium: $U'(r) \to 0$ for $r \to \infty$. In this case, the "jet" velocity in the new reference frame will be less than in the original one: $U'(r) = U'_j = U_j - U_{cf}$ for $r < r_c$. It is easy to show that disturbances $e^{-i\omega(t - x/U_c)}$ in the new coordinate system will have the form $e^{-i\omega'(t - x'/U'_c)}$, where $U'_c = U_c - U_{cf}$, $\omega' = \omega(1 - \kappa)$, $\kappa = U_{cf}/U_c$ (Doppler effect). Further, the dimensionless velocity of the disturbances $\bar{U}_c = U_c/U_j$ in the original coordinate system can be expressed through the dimensionless speed $\bar{U}'_c = U'_c/U'_j$ in the reference frame $(x', r)$

$$\bar{U}_c = \bar{U}'_c(1 - \lambda) + \lambda, \tag{14}$$

where $\lambda = U_{cf}/U_j$ is a co-flow factor, as before. Using the above dependence between frequencies $\omega$ and $\omega'$, one can obtain the following relationship for the Strouhal numbers in both coordinate systems:

$$\mathrm{St} = \mathrm{St}'\left(1 - \lambda + \frac{\lambda}{\bar{U}'_c}\right), \tag{15}$$

where $\mathrm{St}' = \omega' D/(2\pi U'_j)$, $\mathrm{St} = \omega D/(2\pi U_j)$.

Expressions (14) and (15) solve the problem of recalculating the function $\bar{U}'_c(\mathrm{St}')$ obtained for the jet in static conditions ($\lambda = 0$) into the function $\bar{U}_c(\mathrm{St})$ for the jet in flight conditions with the given value of $\lambda$. From the viewpoint of applicability to a real jet, this procedure is, of course, an approximate one and is valid as long as the locally parallel approximation holds true.

Thus, the combination of the model for the convection velocity of the disturbances in a jet under static conditions (8) and the model (14)-(15) for the conversion of this velocity from static to flight conditions makes it possible to recalculate single-point measurements of the velocity fluctuation spectra $L_u$ in the jet near (potential) field into the corresponding pressure spectra $L_p$ in accordance with the relation (see (7))

$$L_p \approx \rho_0^2 (U_c - U_{cf})^2 L_u. \tag{16}$$

An example of implementing the constructed procedure and its validation is described in the next section.

## 3. VALIDATION OF THE PROPOSED METHOD

Let us consider, as an example, a jet of Mach number $M_j = 0.4$ with a co-flow factor $\lambda = 0.23$ and select section $\bar{x} = 3$ to demonstrate the developed method for estimating pressure fluctuations in the co-flow (the bar at the top, as before, denotes nondimensionalization by the nozzle diameter $D$). We consider this flow regime and coordinate $\overline{x}$ because for them, in addition to the near-field hot-wire measurements [34], far-field measurements of the jet-plate interaction noise were obtained. The trailing edge of the plate was located at point $(\overline{x}, \overline{r}) \approx (3,1)$ [36]. This additional far-field data will also be used for the validation of the proposed technique. Application of the method for other values $\overline{x}$ and $\lambda$ is completely similar.

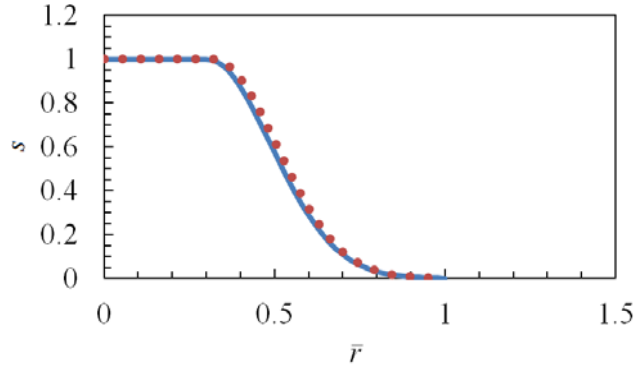

Fig. 7. Comparison of the shapes of the transverse profiles of the average velocity in the corresponding sections for a jet under static conditions (symbols) and a jet in a co-flow $\lambda = 0.23$ (line).

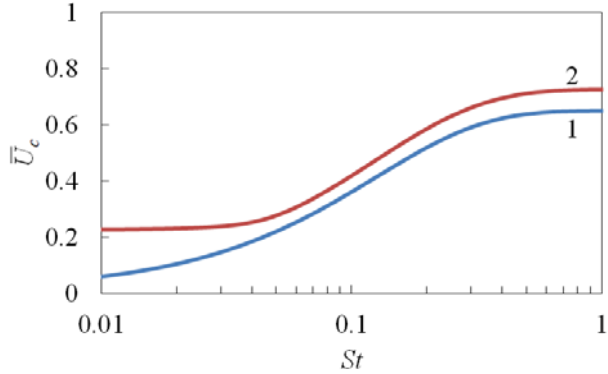

Fig. 8. Relative convection speed of disturbances in the jet at $x / L_c = 0.44$:

$$1 - \lambda = 0; \ 2 - \lambda = 0.23.$$

First, we determine the convection speed of near-field fluctuations for the selected values of $\overline{x}$ and $\lambda$. In accordance with the procedure described above, we consider the jet in static conditions. In this jet, we will search a section $\overline{x}'$ in which the shape of the velocity profile will be equivalent to the one in the original jet at $\overline{x} = 3$. As follows from the mean flow model (11)-(13), the shape of the velocity profiles will be the same at the sections located in a similar way relative to the potential core length. If $\overline{L}_c \approx 6.9$ and $\overline{L}'_c \approx 4.8$ are the lengths of the potential cores of the jets with $\lambda = 0.23$ and $\lambda = 0$, respectively (see (12)), then the velocity profiles will have the same shape in the sections $\bar{x} = 3 \approx 0.44 \bar{L}_c$ and $\bar{x}' = 0.44 \bar{L}'_c \approx 2.1$, respectively. The coincidence of the velocity profiles for the chosen $\bar{x}$ and $\bar{x}'$ can be directly verified using model (11)-(13). This comparison of the profile shapes is shown in Fig. 7, in which, in accordance with (11), a dimensionless function $s = (\bar{U} - \lambda)/(1 - \lambda)$ is plotted for each jet. It is clear that the shapes of the profiles coincide quiet well. It allows (for the given value of $\bar{x}/\bar{L}_c = \bar{x}'/\bar{L}'_c = 0.44$) to estimate the relative convection velocity of the fluctuations in the co-flow using (14), (15), and function (8) modeling the relative convection velocity in static conditions. The corresponding estimate is shown in Fig. 8. As one would expect, in the reference frame associated with the nozzle, the presence of the co-flow generally leads to an increase in the relative convection velocity of the fluctuations due to their transfer by the co-flow. However, this increase is different at different frequencies in accordance with transformation (15) of the frequency scale due to Doppler shift.

Indirect validation of the correctness of the convection velocity estimation can be performed similarly to how it was done in the previous section by scaling the fluctuation spectra along the radial coordinate. For the spectrum of the velocity fluctuations in the section $\bar{x}$, as well as for the pressure fluctuations (see (4), (5), (10)), which are predominantly longitudinal waves in the potential part of the jet near field, the following relation holds true:

$$L_u(\mathrm{St}, \bar{x}, \bar{r}) \propto \left| I_0(\beta_j \bar{r} D) \right|^2, \text{ in the potential core } (\bar{r} < \bar{r}_c), \tag{17}$$

$$L_u(\mathrm{St}, \bar{x}, \bar{r}) \propto \left| K_0(\beta_{cf} \bar{r} D) \right|^2, \text{ outside the mixing layer.} \tag{18}$$

These estimates take into account that inside the potential core, as it approaches the jet axis, only the axisymmetric mode increasingly dominates due to the fact that $I_n(z) \to 0$ for $z \to 0$, $n \geq 1$. In addition, outside the mixing layer, the first two azimuthal modes ($n = 0,1$) with similar characteristics dominate, so functions $I_0$ and $K_0$ can be used to estimate the behavior of the total fluctuations.

The profiles of the dimensionless mean velocity and the root-mean-square values of the fluctuations of its streamwise component, measured by a hot-wire in the section $\bar{x} = 3$, are demonstrated in Fig. 9a. In Fig. 9b, spectral densities of the streamwise velocity component fluctuations in the given cross section are presented for various values of $\bar{r}$ (for convenience, spectra are shown for every fourth point). Dashed lines in Fig. 9a show the approximate boundaries of the mixing layer, within which the most intense fluctuations associated with the turbulent vortex flow are observed. In the potential part of the flow, velocity fluctuations, although significantly weaker than in the mixing layer, are not equal to zero, which is clearly visible in Fig. 9b, where their spectra are shown.

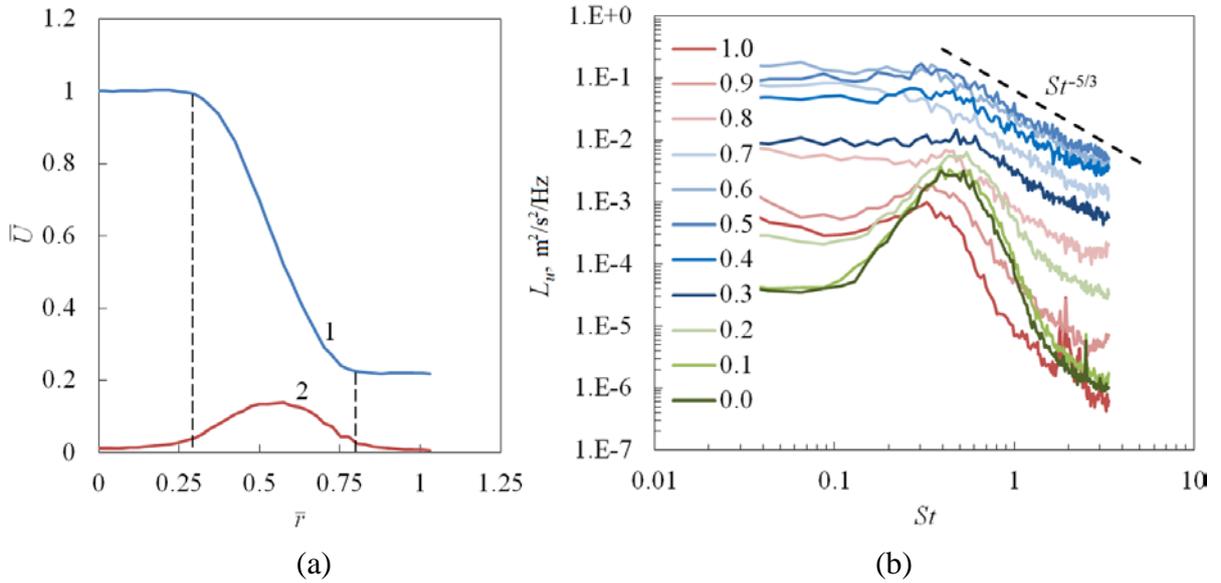

(a)                                                        (b)

Fig. 9. (a) – mean velocity profile (1) and the root-mean-square values of the streamwise velocity (2) in the cross section $\bar{x} = 3$; (b) – spectral densities of the fluctuations of the streamwise velocity component at various values of $\bar{r}$ (curve notation is indicated in the figure).

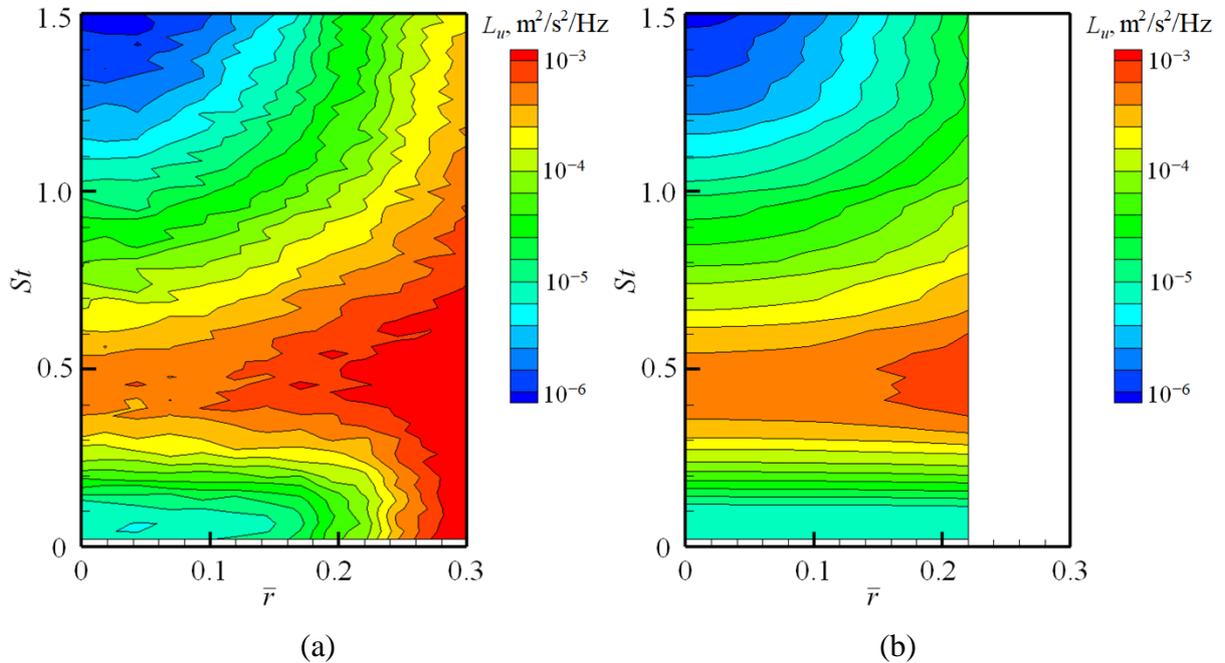

(a)                                                        (b)

Fig. 10. Spectral maps of the velocity fluctuations in the cross section $\bar{x} = 3$ inside the potential core: (a) – experimental data; (b) – modeling result using (17) and the reference spectrum at the point $\bar{r} = 0$.

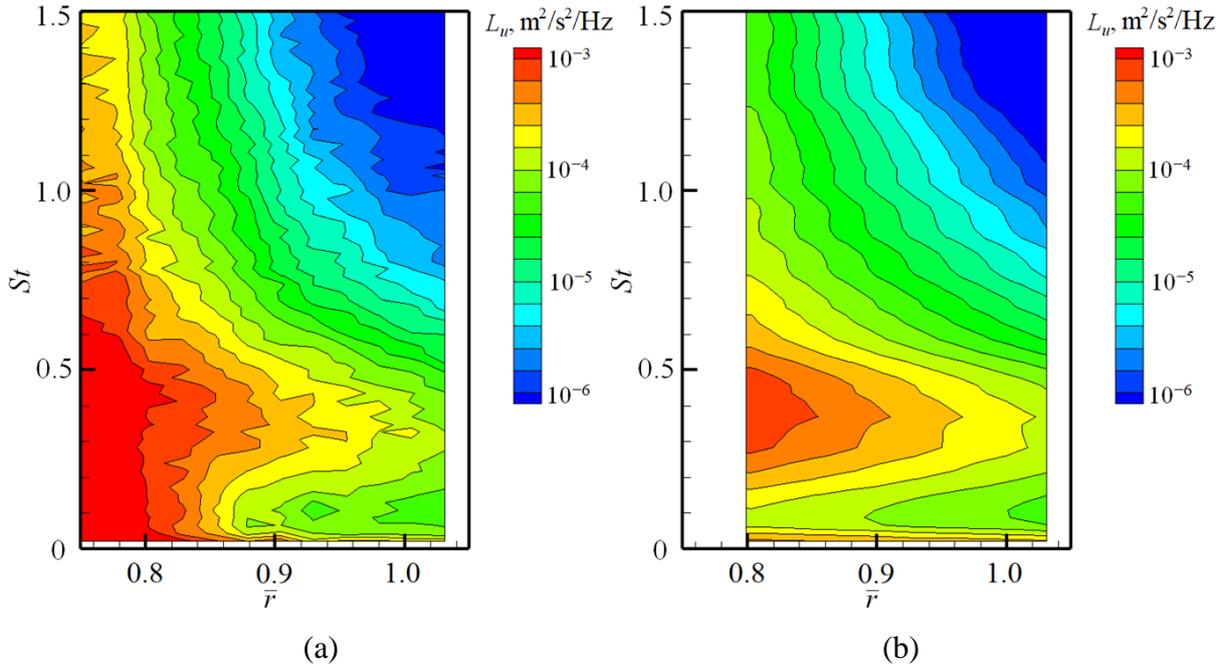

(a)                                              (b)

Fig. 11. Spectral maps of the velocity fluctuations in the cross section $\overline{x} = 3$ in the outer region: (a) – experimental data; (b) – modeling result using (17) and the reference spectrum at the point $\overline{r} = 1$.

From a comparison of the fluctuation spectra at various points in the given section, attention is drawn to the different nature of the spectra inside and outside the mixing layer. In the mixing layer, the spectrum has a shape typical for a developed free turbulent flow with an approximately constant level at low frequencies and a high-frequency decay corresponding the "–5/3" law. In the external region to the mixing layer and in the potential core, the spectrum has a characteristic hump in the range of Strouhal numbers from 0.1 to 1, corresponding to instability waves. In [12, 34], this spectral hump and its connection with instability wave packets were studied in detail using the results of the measurements of the velocity fluctuations on the jet axis. In the presence of the co-flow, as can be seen from Fig. 9b, a hot-wire probe allows measuring wave packet "footprints" outside the jet.

In Fig. 10-11, it is shown that the measurement data in the potential core comply with the model (17), (18) with the convection velocity calculated in accordance with the procedure described above (Fig. 8). In Fig. 10a, the measured spectral map (i.e., the dependence of the fluctuation intensity $L_u$ on $\overline{r}$ and St) of the velocity fluctuations inside the jet is shown. In Fig. 10b, the spectral map constructed on the basis of the reference spectrum $L_u$ at the point $\overline{r} = 0$ using model dependence (17) in the form of $L_u(\mathrm{St}, \overline{r}) \approx L_u(\mathrm{St}, 0)\left|I_0(\beta_j \overline{r} D)\right|^2$ is demonstrated. As can be seen, inside the potential core ($\overline{r} < 0.25$), the model fits the

measurement data quite well. Similarly, Fig. 11a demonstrates the spectral map of the fluctuations measured outside the jet, and Fig. 11b demonstrates the one constructed on the basis of the spectrum $L_u$ at a point $\bar{r} = 1$ using model dependence (18) in the form of $L_u(\mathrm{St}, \bar{r}) \approx L_u(\mathrm{St}, 1)\left|K_0(\beta_{cf}\bar{r}D)\right|^2$. The model fits the measurement data well for the outer near-field region ($\bar{r} > 0.8$). Thus, the proposed semi-empirical model makes it possible to adequately reproduce the spatial-frequency structure of the velocity fluctuations in the near-field potential region both inside and outside the jet. This fact, in turn, justifies the possibility of using expressions (6), (7), (16) for the assessment of the pressure fluctuations in the corresponding regions.

Let us now directly analyze the procedure for estimating the pressure fluctuation spectrum $L_p$ from the velocity fluctuation spectrum $L_u$ in the external region using (16). Direct measurements of the pressure fluctuations in the co-flow were not conducted, because such measurements require, as already noted, positioning of microphones inside the flow and the use of special nose cones [23], which makes them rather cumbersome. However, it is possible to indirectly estimate such fluctuations based on the data on the interaction noise of the jet and a plate placed in its near field. It is shown in [15-17, 35] that if the trailing edge of the plate lies in the potential region of the jet hydrodynamic near-field (Fig. 12), then significant additional noise occurs (the so-called interaction noise). The far-field spectrum of such noise $L_F$ (point F in Fig. 12) is linearly related to the pressure fluctuation spectrum $L_p$ produced by a free jet (in the absence of a plate) at the location of the plate trailing edge (point T in Fig. 12), i.e. $L_F = GL_p$. Thus, the evaluation procedure of $L_p$ proposed in this work based on hot-wire data can be indirectly validated using the acoustic characteristics of the "nozzle-plate" configuration, for which noise measurements in the presence of a co-flow are not difficult, at least at zero angle of attack of the plate [23, 36-38]. In [38], the following simple expression was obtained for the transfer function between the near-field fluctuations and the far-field noise in the symmetry plane of the "nozzle-plate" configuration

$$G \approx \frac{\mathrm{M}_c^3(1+\mathrm{M}_{cf})\sin^2(\theta/2)}{2k^2R^2(1+(\mathrm{M}_{cf}-\mathrm{M}_c))(1+\mathrm{M}_{cf}(\mathrm{M}_c-\mathrm{M}_{cf})-\mathrm{M}_c\cos\theta)^2}, \tag{19}$$

where $\mathrm{M}_{cf} = U_{cf}/c$ and $\mathrm{M}_c = U_c/c$ are Mach numbers of the co-flow and the convective velocity of the near-field perturbations, respectively, $R$ – distance from the trailing edge to the observer, $\theta$ – observation angle (Fig. 12).

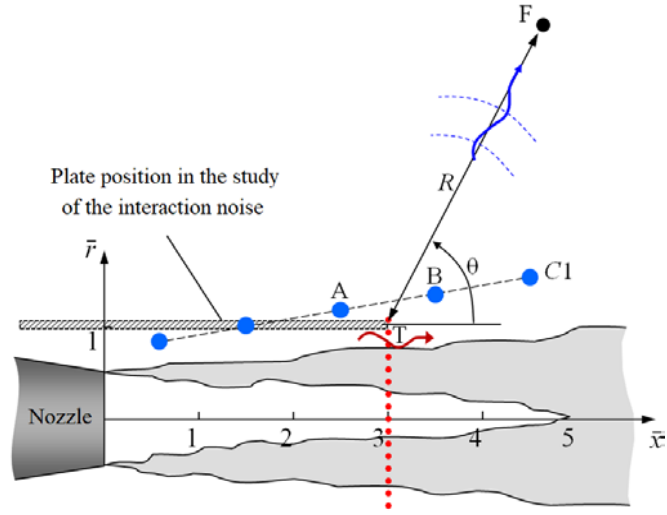

Fig. 12. On the validation of the proposed procedure based on the far-field noise.

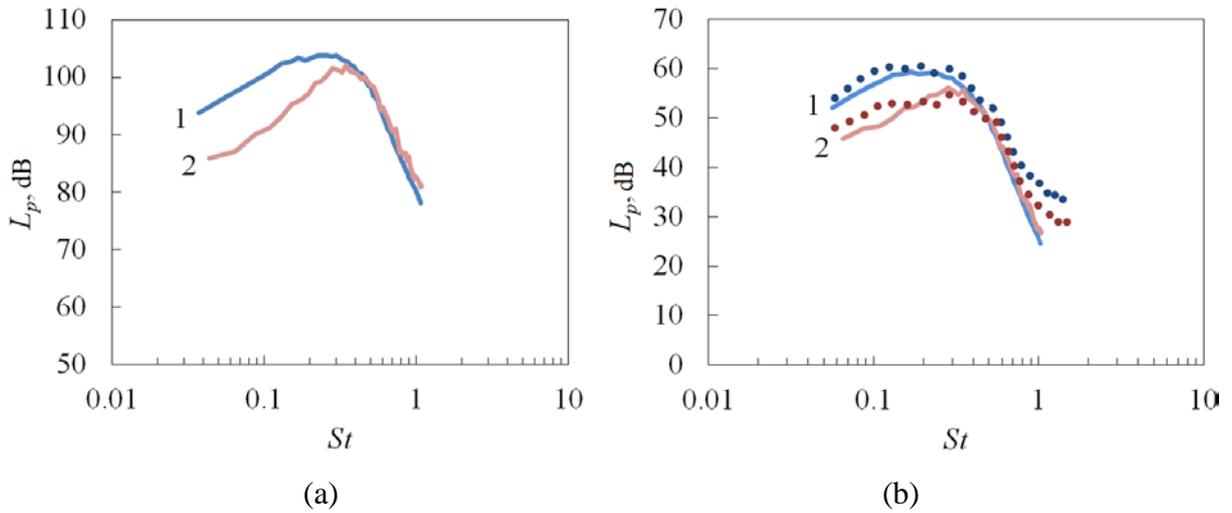

(a)                                                                    (b)

Fig. 13. Spectra of pressure fluctuations in static conditions (1) and in the presence of the co-flow (2): (a) – at the location of the plate trailing edge; (b) – in the far field at the point $R = 20D$, $\theta = -90°$ (point "under" the plate), symbols represent direct measurements of the interaction noise in the far field [36], lines indicate prediction on the basis of the near-field spectra.

Details of the measurements of the jet plate interaction noise, with jet Mach number $M_j = 0.4$ the plate trailing edge located at the point $(\overline{x}, \overline{r}) \approx (3,1)$, in static conditions and in the presence of a co-flow ($\lambda = 0.23$) can be found in [36]. In Fig. 13a, the spectra of the pressure fluctuations in the plate trailing edge location for $\lambda = 0$ and $\lambda = 0.23$ are presented. The spectrum for static conditions was determined from direct measurements of the near field with a microphone array, similar to how it was done in [17]. The spectrum in the presence of the co-flow is calculated using the procedure proposed in this work based on hot-wire measurements of the velocity fluctuations at the point $(\bar{x}, \bar{r}) = (3,1)$. Note that at a fixed coordinate $x$, the presence of the co-flow leads to a slight decrease in the fluctuations level and a shift of their spectrum peak to higher frequencies. The latter effect reflects the fact that the main scale of similarity when comparing different round jets from the viewpoint of the mean flow structure, and therefore the structure of linear perturbations developing on the mean flow, is the potential core length (or the thickness of the mixing layer in a given section linearly related to the potential core length). As the co-flow velocity increases, the length of the potential core increases, and the point with a given coordinate $x$ turns out to be effectively closer to the nozzle exit in terms of the potential core length $L_c$, i.e. the dimensionless parameter $x/L_c$ decreases, which means that the fluctuation spectrum should shift to higher frequencies [39]. A similar effect was observed in [23], where direct pressure fluctuation measurements were carried out using a microphone equipped with a nose cone.

Using the data on the near-field spectra (Fig. 13a) and the transfer function $G$ (19), the low-frequency interaction noise was estimated at a point located under the plate ($\theta = -90°$) at a distance $R = 20D$ from its trailing edge. A comparison of the prediction with the direct measurements of the jet-plate interaction noise is shown in Fig. 13b and demonstrates acceptable accuracy both for static conditions (based on the measurements of the near-field pressure spectra using microphones) and in the presence of the co-flow (based on the hot-wire measurements of the velocity fluctuations). This comparison validates the procedure proposed in the paper for estimating the pressure fluctuations from the velocity ones in the potential near-field region of the jet with co-flow.

In conclusion, we note that the jet-plate interaction noise spectra (and therefore the near-field pressure fluctuations) obtained in this work turn out to be more accurate at low frequencies than those obtained in [36], in which the pressure fluctuations were assessed based on the velocity measurements on the axis jets with their subsequent recalculation to the given point by solving parabolized stability equations. The possibility of direct measurement of the velocity fluctuations outside the jet in the presence of the co-flow, as well as a more correct assessment of their convection velocity based on the experimental data, eliminates the need to use a solution to the stability problem, which is not entirely suitable in the low-frequency region (when the size of the wave packet becomes larger than the length of the potential core of the jet).

# CONCLUSION

The paper proposes a procedure that allows evaluation the characteristics of pressure fluctuations in the jet near-field region external to the mixing layer, in which linear hydrodynamic disturbances (instability waves) dominate, based on the results of single-point hot-wire velocity fluctuation measurements in the presence of a co-flow. Hot-wire measurements in the presence of the co-flow are less invasive and more convenient than microphone measurements. The proposed method is based on the fact that instability waves are close in structure to homogeneous longitudinal waves, which makes it possible to locally relate the pressure fluctuations and the fluctuations of the streamwise velocity component measured by the hot-wire. The key parameter in this relation is the convection velocity of the disturbances. A semi-empirical model for estimating this velocity in the presence of a co-flow is proposed based on the results of its direct measurements with a multi-microphone array in static conditions.

The developed procedure was tested on the experimental data obtained for the Mach 0.4 jet at various co-flow factors. It is shown that velocity fluctuations in the potential region of the near field (both inside the initial part and outside the jet) are well approximated by an axisymmetric solution of the convective wave equation in a locally parallel approximation. The procedure was validated for the far-field jet-plate interaction noise in the presence of the co-flow, and its acceptable accuracy in the range of Strouhal numbers 0.1-1 was demonstrated.

Thus, the developed procedure can be used in a laboratory experiment to reconstruct pressure fluctuation field near the jet in flight conditions in order to assess aeroacoustic loading of airframe elements located near the power plant, as well as additional noise in the far field generated due to the jet-airframe interaction.

## ACKNOWLEDGMENTS

The authors are grateful to V.F. Kopiev for the interest to this work and valuable comments.

## FUNDING

The study was financially supported by the Russian Science Foundation (grant no. 19-71-10064). The experimental part of the study was carried out in TsAGI AC-2 anechoic chamber with flow, upgraded with the financial support of the Ministry of Science and Higher Education of the Russian Federation under agreement no. 075-15-2022-1036.